\documentclass[aps,superscriptaddress,showpacs,nofootinbib,eqsecnum]{revtex4}

\usepackage[dvips]{epsfig}

\begin{document}

\title{Vacuum energies and multipole interactions}

\author{F. A. Barone}
\email{fbarone@unifei.edu.br}
\affiliation{ICE - Universidade Federal de Itajub\'a, Av. BPS 1303, Caixa Postal 50 - 37500-903, Itajub\'a, MG, Brazil.}

\author{G. Flores-Hidalgo}
\email{gflores@cbpf.br}
\affiliation{Departamento de F\'{\i}sica Te\'orica, Universidade do Estado do Rio de Janeiro, 20550-013, Rio de Janeiro, RJ,  Brazil.}

\date{}

\begin{abstract}

In this paper, we present a quantum-field-theoretical description of the interaction between stationary and localized external sources linearly coupled to bosonic fields (specifically, we study models with a scalar and the Maxwell field). We consider external sources that simulate not only point charges but also higher-multipole distributions along D-dimensional branes. Our results complement the ones previously obtained in reference [1].
\end{abstract}

\maketitle

\baselineskip=20pt

\section{Introduction}

In a recent paper \cite{BaroneHidalgo}, we have investigated the role of couplings between quantum fields and external stationary currents (time-independent sources) concentrated along parallel branes with arbitrary co-dimensions. To do that, we have calculated the vacuum energies for a variety of models of quantum fields that interact with external stationary currents concentrated along parallel $D$-branes. As particular cases, for bosonic fields, we have considered external currents which could describe charge distributions and stationary dipole distributions along the branes.

Is is worthy mentioning that systems of quantum fields interacting with external potentials concentrated along branes have been treated in the literature; see, for instance, \cite{bordag,milton,S} and references cited therein. But, the coupling of quantum fields to external currents concentrated along branes is not a well-explored subject.

In order to fulfill a question left aside in reference \cite{BaroneHidalgo}, in this paper we make a deeper discussion on the description of multipole density distributions along branes with arbitrary dimensions by the use of external currents concentrated at specific regions of space. The results for charges and dipole distributions, which can be taken as generalizations of the ones exposed in reference \cite{BaroneHidalgo}, are presented, for completeness, in this paper and also in order to make clearer some discussions and correct a flaw in reference \cite{BaroneHidalgo}. The results obtained here for currents describing four-pole distributions and $N$-pole distributions are novel.

Along the paper, we shall consider models in $d+D+1$ dimensions described by a quantum field coupled to an external current concentrated along a $D$-brane and another one concentrated at a given point of space. The latter current represents a point-like test-charge which is used to investigate the force field produced by the former one. 	
	
	We shall also use the same notation as in reference \cite{BaroneHidalgo}, where the coordinate $(D+d+1)$-vector is given by
\begin{equation}
\label{def4vetor}
x=(x^{0},x^{1},...,x^{d},x^{d+1},...,x^{d+D})\ ,
\end{equation}
and its perpendicular and parallel parts to the $D$-brane are, respectively,
\begin{eqnarray}
\label{defxperpx|}
{\bf x}_{\perp}&=&(x^{1},...,x^{d})\ ,\nonumber\\
{\bf x}_{\|}&=&(x^{d+1},...,x^{d+D})\ .
\end{eqnarray}
We shall also use similar notations for the momenta $k$, as well as for any other vector considered in this paper.

	This work is outlined as follows: in section (\ref{escalar}), we consider models for the scalar field, with and without mass, coupled to currents describing distributions of charges, dipoles, quadrupoles and $N$-poles. Section (\ref{eletromagnetico}) is devoted to extending the previous results for the electromagnetic case. In section (\ref{conclusao}), we draw some final remarks and conclusions.
	
\section{Scalar Field}
\setcounter{equation}{0}
\label{escalar}

	All over this section, we consider models for the massive scalar field, $\phi$, in $d+D+1$ dimensions, always interacting with an external current $J$ different for each model.	The lagrangians of all models investigated have the same structure
\begin{equation}
\label{Lescalar}
{\cal L}_{scalar}=\frac{1}{2}(\partial_{\mu}\phi)(\partial^{\mu}\phi)-\frac{1}{2}m^{2}\phi^{2}+J\phi\ ,
\end{equation}
with the corresponding generating functional of the Green's functions
\begin{equation}
\label{rfv}
{\cal Z}=\exp\Biggl(-\frac{i}{2}\int\int\ d^{d+D+1}x\ d^{d+D+1}y\ \ J(x)\Delta_{F}(x,y)J(y)\Biggr)\ ,
\end{equation}
where $\Delta_{F}(x,y)$ is the Green's function
\begin{equation}
\label{defDeltaF}
\Delta_{F}(x,y)=\lim_{\varepsilon\to0}\int{\frac{d^{d+D+1}k}{(2\pi)^{d+D+1}}\ \frac{\exp\Bigl[ik(x-y)\Bigr]}{k^{2}-m^{2}+i\varepsilon}}\ .
\end{equation}

	It is worthy mentioning that the current $J$ in the functional (\ref{rfv}) is not an auxiliary parameter introduced in order to perform perturbative calculations, as usualy considered in the literature. The current  $J$ has here a physical meaning and shall not be set equal to zero.
	
	In the limit $T\to\infty$, the generating functional of any quantum system, whose lagrangian density does not depend explicitly on the time coordinate (what is the case we are considering), can be written in the form \cite{Zee,Itzykson,Peskin}
\begin{equation}
\label{FG}
{\cal Z}=\exp(-iET)\ ,
\end{equation}
where $E$ is the lowest energy of the system and $T=\int_{-\infty}^{\infty}dx^{0}$.

	Comparing Eq's (\ref{rfv}) and (\ref{FG}), we have
\begin{equation}
\label{rfv1}
E=\lim_{T\rightarrow 0}\frac{1}{2T}\int\int\ d^{d+D+1}x\ d^{d+D+1}y\ \ J(x)\Delta_{F}(x,y)J(y)\ .
\end{equation}

	From now on, let us take the current $J$ to be composed by a fuction concentrated along a $D$-dimensional brane along with another function concentrated at a given point of space.

	The first model we study is a generalization of the one exposed in \cite{BaroneHidalgo} and it is considered in this work for completeness and in order to make clearer the method employed in this whole paper in a simple example. The current corresponding to the first model is taken to be given by
\begin{equation}
\label{corrente1}
J_{I}({\bf x})=\sigma\delta^{d}({\bf x}_{\perp}-{\bf A})+\sigma_{0}\delta^{d+D}({\bf x}-{\bf a})\ ,
\end{equation}
where ${\bf A}=(A^{1},A^{2},...,A^{d})$ and ${\bf a}=(a^{1},a^{2},...,a^{d+D})$. The first term at the right-hand side of (\ref{corrente1}) is a distribution concentrated along a $D$-dimensional brane denoted by the vector ${\bf A}$, the last term at the right-hand side is a distribution concentrated at the point ${\bf a}$.

	Substituting (\ref{defDeltaF}) and (\ref{corrente1}) into (\ref{rfv1}), discarding terms due to the brane self-interaction and the test-charge self-interaction, performing, in the following order, the integrals $dx^{0}dk^{0}dy^{0}d^{D}{\bf x}_{\|}d^{D}{\bf y}_{\|}d^{D}{\bf k}_{\|}d^{d}{\bf x}_{\perp}d^{d}{\bf y}_{\perp}$ and making a trivial change of variables, we arrive at
\begin{equation}
\label{qwe1}
E_{I}=-\sigma\sigma_{0}\int\frac{d^{d}{\bf k}_{\perp}}{(2\pi)^{d}}\frac{\exp[-i{\bf k}_{\perp}\cdot({\bf a}_{\perp}-{\bf A})]}{{\bf k}_{\perp}^{2}+m^{2}}\ ,
\end{equation}
where we have used that $T=\int dx^{0}$.

	The analytic extension for the integral in (\ref{qwe1}) is calculated in reference \cite{BaroneHidalgo}, where we consider, separately, the situations with and without mass. For $m=0$ and $d\not=2$, we have
\begin{eqnarray}
\label{qwe2}
E_{I}(m&=&0,d\not=2)\nonumber \\
&=&-\frac{\sigma\sigma_{0}}{(2\pi)^{d/2}}2^{(d/2)-2}\Gamma\Bigl((d/2)-1\Bigr)|{\bf a}_{\perp}-{\bf A}|^{2-d}\ ,\cr
&\ &
\end{eqnarray}
where $\Gamma$ is the gamma function.

	For a massive field, we have
\begin{eqnarray}
\label{qwe3}
&&E_{I}(m,d)=-\frac{\sigma\sigma_{0}}{(2\pi)^{d/2}}m^{d-2}\nonumber \\
&&\Bigl(m|{\bf a}_{\perp}-{\bf A}|\Bigr)^{1-(d/2)}K_{(d/2)-1}(m|{\bf a}_{\perp}-{\bf A}|),\
\end{eqnarray}
where $K_{\mu}(x)$ designates the $K$-Bessel function \cite{Arfken}. It is worth mentioning that expression (\ref{qwe3}) is valid for any $d>0$, even for $d=2$.

	Assuming that $d\not=2$ and taking the limit $m\to 0$ in Eq. (\ref{qwe3}), we obtain the result (\ref{qwe2}), what can be done with the aid of the expression $K_{\nu}(z)\stackrel{z\to0}{\longrightarrow}\Gamma(\nu)2^{\nu-1}/z^{\nu}\ ,\ \nu\not=0$ . The energy for the situation where $m=0$ and $d=2$ is obtained with the help of (\ref{qwe3}) and the expression $K_{0}(z)\stackrel{z\to 0}{\longrightarrow}-\ln(z/2)-\gamma$ \cite{Arfken}, with $\gamma$ designating the Euler constant, as follows
\begin{eqnarray}
\label{qwe4}
E_{I}(m=0,d=2)&=&-\frac{\sigma\sigma_{0}}{2\pi}\lim_{m\to 0}\Biggl[K_{0}(m|{\bf a}_{\perp}-{\bf A}|)\Biggr]\cr\cr
&\cong&-\frac{\sigma\sigma_{0}}{2\pi}\lim_{m\to 0}\Biggl[-\ln\Biggl(\frac{m|{\bf a}_{\perp}-{\bf A}|}{2}\Biggr)-\gamma\Biggr]\cr\cr
&\cong&-\frac{\sigma\sigma_{0}}{2\pi}\lim_{m\to 0}\Biggl[-\ln\Biggl(\frac{m|{\bf a}_{\perp}-{\bf A}|}{2}\Biggr)-\gamma\cr\cr
&\ &+\ln(ma_{0})-\ln(ma_{0})\Biggr]\cr\cr
&\cong&\frac{\sigma\sigma_{0}}{2\pi}\ln\Biggl(\frac{|{\bf a}_{\perp}-{\bf A}|}{a_{0}}\Biggr)\cr\cr
&\ &+\frac{\sigma\sigma_{0}}{2\pi}\Bigl[\gamma-\ln(2)+\lim_{m\to 0}\ln(ma_{0})\Bigr]\cr\cr
&\rightarrow& \frac{\sigma\sigma_{0}}{2\pi}\ln\Biggl(\frac{|{\bf a}_{\perp}-{\bf A}|}{a_{0}}\Biggr)\ ,
\end{eqnarray}
where, in the fourth line, we have added and subtracted the term $\ln(ma_{0})$ inside the brackets, introducing an arbitrary length-dimensional finite constant $a_{0}$.
In the last line of the above expression, we have discarded all terms which do not depend on the distance $|{\bf a}_{\perp}-{\bf A}|$, even the divergent ones, once they do not contribute to the force between the test charge and the brane.

The presence of the arbitrary constant $a_{0}$ in the energy (\ref{qwe4}) does not produce any physical result, once the force between the test charge and the brane do not depend on the distance $|{\bf a}_{\perp}-{\bf A}|$. In fact, one could add the constant term $(\sigma\sigma_{0})/(2\pi)\ln(a_{0})$ to the energy (\ref{qwe4}), what leads to $E_{I}(m=0,d=2)=(\sigma\sigma_{0})/(2\pi)\ln(|{\bf a}_{\perp}-{\bf A}|)$. The constant $a_{0}$ was introduced for convenience, in order to make the argument of the logarithm dimensionless.

	Let us take the restriction $D+d=3$, which corresponds to adopting a space-time with $3+1$ dimensions. In this case, we have, for $d=1$, $d=2$ and $d=3$, respectively, a uniform distribution of charges along a plane, a straight line and a point. The energy corresponding to the masless case and $d=2$ is given by (\ref{qwe4}). For the masless case, with $d=1$ and $d=3$, the results for the energy (\ref{qwe2}) read, respectively,
\begin{eqnarray}
\label{qwe5}
E_{I}(m=0,d=1)&=&\frac{\sigma\sigma_{0}}{2}|{\bf a}_{\perp}-{\bf A}|\cr\cr
E_{I}(m=0,d=3)&=&\frac{\sigma\sigma_{0}}{4\pi}|{\bf a}-{\bf A}|^{-1}\ ,
\end{eqnarray}
where, in the last line, we have suppressed the sub-index $\perp$ for the vector ${\bf a}$, once its parallel part does not exist whenever $D=0$, what happens once $D+d=3$ and $d=3$.

	The results (\ref{qwe4}) and (\ref{qwe5}) agree with the ones obtained for the electromagnetic field in classical electrodynamics up to an overall sign, always present in comparing the scalar and electromagnetic fields. The second Eq. (\ref{qwe5}) is the coulombian interaction between two point charges $\sigma_{0}$ and $\sigma$ placed at positions ${\bf a}$ and ${\bf A}$, respectively, obtained in references \cite{Zee,BaroneHidalgo,Itzykson}.

	The second current we study is given by
\begin{equation}
\label{corrente2}
J_{II}({\bf x})=\sigma V^{\mu}\Bigl(\partial_{\mu}\delta^{d}({\bf x}_{\perp}-{\bf A})\Bigr)+\sigma_{0}\delta^{d+D}({\bf x}-{\bf a})\ ,
\end{equation}
where $V^{\mu}$ is a four-vector taken to be constant and uniform in the reference frame we are performing the calculations, and also, with vanishing time and perpendicular components in this frame, that is, $V^{0}=0$ and ${\bf V}_{\|}=0$. The partial derivative in the above equation is with respect to the ${\bf x}$ coordinates.

	Substituting (\ref{corrente2}) in (\ref{rfv1}), discarding terms due to self interactions, as before, performing a change of integration variables and an integration by parts we have
\begin{eqnarray}
\label{qwe6}
E_{II}&=&-\frac{\sigma\sigma_{0}}{T}\int\int d^{d+D+1}x\ d^{d+D+1}y\  \delta^{d}({\bf x}_{\perp}-{\bf A})\cr\cr
&\ &\times\ \delta^{d+D}({\bf y}-{\bf a})({\bf V}\cdot{\bf\nabla}_{\perp})\Delta_{F}(x,y)\ ,
\end{eqnarray}
where we have defined the operator ${\bf\nabla}_{\perp}=(\partial/\partial x^{1},\partial/\partial x^{2},...,\partial/\partial x^{d})$.

	Using the Fourier representatin (\ref{defDeltaF}) in (\ref{qwe6}), operating with $({\bf V}\cdot{\bf \nabla}_{\perp})$, performing, in the following order, the integrals $dx^{0}dk^{0}dy^{0}d^{D}{\bf x}_{\|}d^{D}{\bf y}_{\|}d^{D}{\bf k}_{\|}d^{d}{\bf x}_{\perp}d^{d}{\bf y}_{\perp}$ and using the fact that $T=\int dx^{0}$ we have
\begin{eqnarray}
\label{azxc1}
E_{II}&=&-\sigma\sigma_{0}\int\frac{d^{d}{\bf k}_{\perp}}{(2\pi)^{d}}\frac{{\bf V}_{\perp}\cdot i{\bf k}_{\perp}}{{\bf k}_{\perp}^2+m^{2}}\exp\Bigl[i{\bf k}_{\perp}\cdot({\bf a}_{\perp}-{\bf A})\Bigr]\cr\cr
&=&-\sigma\sigma_{0}\bigl({\bf V}_{\perp}\cdot{\bf \nabla}_{a\perp}\bigr)\int\frac{d^{d}{\bf k}_{\perp}}{(2\pi)^{d}}\frac{1}{{\bf k}_{\perp}^2+m^{2}}\cr\cr
&\ &\exp\Bigl[i{\bf k}_{\perp}\cdot({\bf a}_{\perp}-{\bf A})\Bigr] ,
\end{eqnarray}
where we have defined the differential operator ${\bf\nabla}_{a\perp}=(\partial/\partial a^{1},...,\partial/\partial a^{d})$.

	The integral which appears in (\ref{azxc1}) is the same one present in (\ref{qwe1}) up to the sign of the exponential argument. This different sign is irrelevant for the result \footnote{With a change of variables this sign can be inverted} . As already said, this integral is calculated in reference \cite{BaroneHidalgo} for $m=0$ and $m\not=0$, separately.
	
	By using (\ref{qwe1}) and (\ref{qwe2}), the energy (\ref{azxc1}) for the masless case reads
\begin{eqnarray}
\label{azxc2}
E_{II}(m=0,d)&=&-\sigma\sigma_{0}\bigl({\bf V}_{\perp}\cdot{\bf \nabla}_{a\perp}\bigr)\frac{1}{(2\pi)^{d/2}}2^{(d/2)-2}\cr\cr
&\ &\Gamma\Bigl((d/2)-1\Bigr)|{\bf a}_{\perp}-{\bf A}|^{2-d}\cr\cr
&=&-\frac{\sigma_{0}}{(2\pi)^{d/2}}2^{(d/2)-1}\ \Gamma\bigl(d/2\bigr)\cr\cr
&\ &|{\bf a}_{\perp}-{\bf A}|^{1-d}\ (-\sigma{\bf V})\cdot\frac{{\bf a}_{\perp}-{\bf A}}{|{\bf a}_{\perp}-{\bf A}|} .
\end{eqnarray}

	As we shall see, result (\ref{azxc2}) is the interaction energy between a point charge $\sigma_{0}$ at the position ${\bf a}$ and a dipole distribution along a $D$-brane placed at ${\bf A}$ with dipole density $-\sigma{\bf V}$.
		
	Comparison of Eq's (\ref{qwe1}) and (\ref{qwe3}) allows us to write down the energy (\ref{azxc1}), for the massive field, in the form
\begin{eqnarray}
\label{azxc3}
E_{II}(m,d)&=&-\frac{\sigma\sigma_{0}}{(2\pi)^{d/2}}m^{d-2}\bigl({\bf V}_{\perp}\cdot{\bf \nabla}_{a\perp}\bigr)\cr\cr
&\ &\Bigl[\Bigl(m|{\bf a}_{\perp}-{\bf A}|\Bigr)^{1-(d/2)}K_{(d/2)-1}\Bigl(m|{\bf a}_{\perp}-{\bf A}|\Bigr)\Bigr]\cr\cr
&=&\frac{\sigma_{0}}{(2\pi)^{d/2}}m^{d/2}|{\bf a}_{\perp}-{\bf A}|^{-d/2}\cr\cr
&\ &K_{d/2}\Bigl(m|{\bf a}_{\perp}-{\bf A}|\Bigr)\Bigl((-\sigma{\bf V})\cdot({\bf a}_{\perp}-{\bf A})\Bigl) ,
\end{eqnarray}
where, in the last line, we used the fact that
\begin{equation}
\frac{d}{dx}\Bigl(x^{1-(d/2)}K_{(d/2)-1}(x)\Bigr)=-x^{1-(d/2)}K_{d/2}(x)\ .
\end{equation}
Taking the limit of vanishing mass in the result (\ref{azxc3}), we obtain the energy (\ref{azxc2}), even for the case $d=2$.

	Now, we restrict to the case $D+d=3$,  which means that we are in a $3+1$ space-time. In this case, taking $d=1$, $d=2$ and $d=3$ means that the brane is reduced to a plane, a line and a point, respectively, and the corresponding energies read
\begin{eqnarray}
\label{azxc4}
E_{II}(m,d=1)&=&\frac{\sigma\sigma_{0}}{2}|{\bf a}_{\perp}-{\bf A}|^{-1}\cr\cr
&\ &\exp{(m|{\bf a}_{\perp}-{\bf A}|)}{\bf V}\cdot({\bf a}_{\perp}-{\bf A})\ ,\cr\cr
E_{II}(m,d=2)&=&\frac{\sigma\sigma_{0}}{2\pi}m|{\bf a}_{\perp}-{\bf A}|^{-1}\cr\cr
&\ &K_{1}(m|{\bf a}_{\perp}-{\bf A}|){\bf V}\cdot({\bf a}_{\perp}-{\bf A})\ ,\cr\cr
E_{II}(m,d=3)&=&\frac{\sigma\sigma_{0}}{4\pi}m|{\bf a}-{\bf A}|^{-2}\exp{(-m|{\bf a}-{\bf A}|)}\cr\cr
&\ &\Biggl(1+\frac{1}{m|{\bf a}-{\bf A}|}\Biggr){\bf V}\cdot({\bf a}-{\bf A})\ ,
\end{eqnarray}
where, in the last line, we have discarded the sub-index $\perp$, once $D=0$.

	For a vanishing mass, the last equation (\ref{azxc4}) reads
\begin{equation}
\label{dipolospontuais}
E_{II}(m=0,d=3)=-\frac{\sigma_{0}\sigma}{4\pi}\frac{(-\sigma{\bf V})\cdot({\bf a}-{\bf A})}{|{\bf a}-{\bf A}|^{-2}}\ .
\end{equation}
which is the interaction energy between a test scalar charge $\sigma$ placed at the point ${\bf a}$ and a scalar dipole $-\sigma{\bf V}$ lying in the position ${\bf A}$. It is important to notice that, in comparing with the electromagnetic field, we have an overall minus sign.

	From the above computations, we can interpret the first term in the current (\ref{corrente2}) as a uniform distribution of stationary dipoles with dipole momentum density given by $-\sigma{\bf V}$ along a $D$-dimensional brane placed at ${\bf A}$.

	In the paper \cite{BaroneHidalgo}, we have considered a scalar current composed by an arbitrary number $N$ of terms similar to the first one present  in (\ref{corrente2}). Each term was taken to be concentrated along a different brane and all of them were taken to be parallel to one another. We have also taken a different four-vector $V^{\mu}_{(i)}$ ($i=1..N$) for each term. In order to analyze the meaning of the considered current, we have considered two point-like branes and a $3+1$ space-time. Once the interaction energy we have obtained contains only terms proportional to the products ${\bf V}_{i}\cdot{\bf V}_{j}$, ($i=1,2$), as shown in equation (43) of reference \cite{BaroneHidalgo}, that is
\begin{equation}
{\cal E}=\frac{-1}{4\pi a^{3}}\Bigl[\Bigr(\sigma_{1}{\bf V}_{(1)}\cdot\sigma_{2}{\bf V}_{(2)}\Bigr)-3\Bigl(\sigma_{1}{\bf V}_{(1)}\cdot{\hat a}\Bigr)\!\!\!\Bigl(\sigma_{2}{\bf V}_{(2)}\cdot{\hat a}\Bigr)\Bigr]\ ,
\end{equation}
we have interpreted each term in the current as a dipole with the wrong sign. In this paper, we correct this point and the true interpretation is the one exposed after Eq. (\ref{dipolospontuais}).

	The third and last model we consider for the scalar field is determined by the current
\begin{equation}
\label{corrente3}
J_{III}({\bf x})=\sigma V^{\mu\nu}\Bigl(\partial_{\mu}\partial_{\nu}\delta^{d}({\bf x}_{\perp}-{\bf A})\Bigr)+\sigma_{0}\delta^{d+D}({\bf x}-{\bf a})\ , 	
\end{equation}
where $V^{\mu\nu}$ is a symmetric tensor with rank-2 taken to be constant and uniform in the refrence frame we are performing the calculations. From (\ref{corrente3}), it can be noticed that we can take $V^{0\mu}=V^{i\mu}=0$, $i=d+1,...,D$ with no loss of generality.

	Substituting the current (\ref{corrente3}) in equation (\ref{rfv1}), performing two integrations by parts, integrating in the variables $dx^{0}dk^{0}dy^{0}d^{D}{\bf x}_{\|}d^{D}{\bf y}_{\|}d^{D}{\bf k}_{\|}d^{d}{\bf x}_{\perp}d^{d}{\bf y}_{\perp}$ and using the definitions of $T$ and the operator ${\bf\nabla}_{a\perp}$, both employed in equation (\ref{azxc1}), we have
\begin{eqnarray}
\label{azxc5}
&&E_{III}(m,d)=\cr\cr
&&\!\!\!\!\!\!\!\!\!\! =-\sigma\sigma_{0}\sum_{i,j=1}^{d}\int\frac{d^{d}{\bf k}_{\perp}}{(2\pi)^{d}}\frac{(i{\bf k}_{\perp})^{i}\ V^{ij}\ (i{\bf k}_{\perp})^{j}}{{\bf k}_{\perp}^{2}+m^{2}}\cr\cr
&& \exp{\bigl[i{\bf k}_{\perp}\cdot({\bf a}_{\perp}-{\bf A})\bigr]}\cr\cr
&&\!\!\!\!\!\!\!\!\!\! =-\sigma\sigma_{0}\sum_{i,j=1}^{d}V^{ij}{\bf\nabla}_{a\perp}^{i}{\bf\nabla}_{a\perp}^{j}\int\frac{d^{d}{\bf k}_{\perp}}{(2\pi)^{d}}\frac{1}{{\bf k}_{\perp}^{2}+m^{2}}\cr\cr
&&\exp{\bigl[i{\bf k}_{\perp}\cdot({\bf a}_{\perp}-{\bf A})\bigr]} .
\end{eqnarray}

	As before, we first consider the masless case and, next, the massive field. For this model, this two-step analysis is important in order to identify a freedom in the choice of $V^{\mu\nu}$.
	
	Once we do not take the test charge in the brane, that is, ${\bf a}_{\perp}\not={\bf A}$, and using Fourier representation for the Dirac delta function, we can write
\begin{eqnarray}
\label{azxc6}
&&\!\!\!\!\!\!\!\!\!\!\!\!\!\!\!\!\!\!\!\!\!\!\!\!\!\!\!\!\!\! {\bf\nabla}_{a\perp}^{2}\int\frac{d^{d}{\bf k}_{\perp}}{(2\pi)^{d}}\frac{1}{{\bf k}_{\perp}^{2}}\exp{\bigl[i{\bf k}_{\perp}\cdot({\bf a}_{\perp}-{\bf A})\bigr]}=\nonumber \\
&&\!\!\!\!\!\!\!\!\!\!\!\!\!\!\!\!\!\!\!\!\!\!\!\!\!\!\!\!\!\! -\int\frac{d^{d}{\bf k}_{\perp}}{(2\pi)^{d}}\exp{\bigl[i{\bf k}_{\perp}\cdot({\bf a}_{\perp}-{\bf A})\bigr]}=\delta^{d}({\bf a}_{\perp}-{\bf A})=0\ .
\end{eqnarray}

	For the masless case, the energy (\ref{azxc5}) reads
\begin{eqnarray}
\label{azxc7}
&&E_{III}(m=0,d)=\cr\cr
&&-\sigma\sigma_{0}\sum_{i,j=1}^{d}V^{ij}{\bf\nabla}_{a\perp}^{i}{\bf\nabla}_{a\perp}^{j}\int\frac{d^{d}{\bf k}_{\perp}}{(2\pi)^{d}}\frac{1}{{\bf k}_{\perp}^{2}}\exp{\bigl[i{\bf k}_{\perp}\cdot({\bf a}_{\perp}-{\bf A})\bigr]}\cr\cr
&&+\sigma\sigma_{0}\frac{1}{d}(tr V)\left[{\bf\nabla}_{a\perp}^{2}\int\frac{d^{d}{\bf k}_{\perp}}{(2\pi)^{d}}\frac{1}{{\bf k}_{\perp}^{2}}\exp{\bigl[i{\bf k}_{\perp}\cdot({\bf a}_{\perp}-{\bf A})\bigr]}\right]\ ,\cr
&&\ 
\end{eqnarray}
where we have introduced a vanishing term (the second one in the right hand side), as stated in expression (\ref{azxc6}). In the above equation, $tr V$ stands for the trace of tensor $V$.

	Defining the traceless tensor
\begin{equation}
\label{defD}
D^{ij}=V^{ij}-\frac{tr V}{d}\delta^{ij}\ ,
\end{equation}
the energy (\ref{azxc7}) can be rewritten in the form
\begin{eqnarray}
\label{aqwe1}
&&E_{III}(m=0,d)=\nonumber \\
&&-\sigma\sigma_{0}\sum_{i,j=1}^{d}D^{ij}{\bf\nabla}_{a\perp}^{i}{\bf\nabla}_{a\perp}^{j}\int\frac{d^{d}{\bf k}_{\perp}}{(2\pi)^{d}}\frac{1}{{\bf k}_{\perp}^{2}}\exp{\bigl[i{\bf k}_{\perp}\cdot({\bf a}_{\perp}-{\bf A})\bigr]}\ .\cr
&&
\end{eqnarray}

	If we compare Eq's (\ref{qwe1}) and (\ref{qwe2}), we can obtain the integral which appears in the above equation and write the energy (\ref{aqwe1}) in the form
\begin{eqnarray}
\label{aqwe2}
E_{III}(m=0,d)=
-\frac{\sigma\sigma_{0}}{(2\pi)^{d/2}}2^{(d/2)-2}\Gamma\bigl((d/2)-1\bigr)\cr\cr
\sum_{i,j=1}^{d}D^{ij}{\bf\nabla}_{a\perp}^{i}{\bf\nabla}_{a\perp}^{j}|{\bf a}_{\perp}-{\bf A}|^{2-d}\cr\cr
=-\frac{\sigma_{0}}{\pi^{d/2}}\Gamma\bigl((d/2)+1\bigr)\frac{1}{|{\bf a}_{\perp}-{\bf A}|^{d}}\cr\cr
\sum_{i,j=1}^{d}\frac{({\bf a}_{\perp}-{\bf A})^{i}}{|{\bf a}_{\perp}-{\bf A}|}(\sigma D^{ij})\frac{({\bf a}_{\perp}-{\bf A})^{j}}{|{\bf a}_{\perp}-{\bf A}|}\ .
\end{eqnarray}
	
	The result (\ref{aqwe2}) is the interaction energy between a point charge and a four-pole distribution along a $D$-brane with four-pole momentum desnsity given by $\sigma D^{ij}$. This point can be made clearer if we consider a $3+1$ space-time, which corresponds to take $D+d=3$. With this restriction $d$ can assume the values $1$, $2$ and $3$, leading to the energies
\begin{widetext}
\begin{eqnarray}
\label{aqwe3}
E_{III}(m=0,d=1)&=&-\frac{\sigma_{0}}{2}\frac{1}{|{\bf a}_{\perp}-{\bf A}|}\frac{({\bf a}_{\perp}-{\bf A})^{i}}{|{\bf a}_{\perp}-{\bf A}|}(\sigma D^{ij})\frac{({\bf a}_{\perp}-{\bf A})^{j}}{|{\bf a}_{\perp}-{\bf A}|}\ ,\cr\cr
E_{III}(m=0,d=2)&=&-\frac{\sigma_{0}}{\pi}\frac{1}{|{\bf a}_{\perp}-{\bf A}|^{2}}\frac{({\bf a}_{\perp}-{\bf A})^{i}}{|{\bf a}_{\perp}-{\bf A}|}(\sigma D^{ij})\frac{({\bf a}_{\perp}-{\bf A})^{j}}{|{\bf a}_{\perp}-{\bf A}|}\ ,\cr\cr
E_{III}(m=0,d=3)&=&-\frac{3\sigma_{0}}{4\pi}\frac{1}{|{\bf a}-{\bf A}|^{3}}\frac{({\bf a}-{\bf A})^{i}}{|{\bf a}-{\bf A}|}(\sigma D^{ij})\frac{({\bf a}-{\bf A})^{j}}{|{\bf a}-{\bf A}|}\ .	
\end{eqnarray}
\end{widetext}

	The last equation ({\ref{aqwe3}) is the interaction energy between a point charge placed at ${\bf a}$ and a point-like four-pole placed at ${\bf A}$ with four pole momentum given by $\sigma D^{ij}$.

	It is interesting to notice that the trace of the tensor $V$ in (\ref{corrente3}) is irrelevant for the energy (\ref{aqwe2}). This situation is different when we consider the field with mass, as it shall be shown.
	
	When $m\not=0$, we can compare Eq's (\ref{qwe1}) and (\ref{qwe3}) in order to obtain the integral of Eq. (\ref{azxc5}), which takes the form
\begin{eqnarray}
\label{aqwe4}
&&E_{III}(m,d)=-\frac{\sigma\sigma_{0}}{(2\pi)^{d/2}}m^{d-2}\sum_{i,j=1}^{d}V^{ij}{\bf\nabla}_{a\perp}^{i}{\bf\nabla}_{a\perp}^{j}\cr\cr
&&\Bigl[(m|{\bf a}_{\perp}-{\bf A}|)^{1-(d/2)}K_{(d/2)-1}(m|{\bf a}_{\perp}-{\bf A}|)\Bigr]\cr\cr
&&=\frac{\sigma_{0}}{(2\pi)^{d/2}}m^{d/2}|{\bf a}_{\perp}-{\bf A}|^{-d/2}\cr\cr
&&\sum_{i,j=1}^{d}(\sigma V^{ij})\Biggl[\delta^{ij}K_{d/2}(m|{\bf a}_{\perp}-{\bf A}|)\cr\cr
&&-m|{\bf a}_{\perp}-{\bf A}|\frac{({\bf a}_{\perp}-{\bf A})^{i}}{|{\bf a}_{\perp}-{\bf A}|}\frac{({\bf a}_{\perp}-{\bf A})^{j}}{|{\bf a}_{\perp}-{\bf A}|}K_{(d/2)+1}(m|{\bf a}_{\perp}-{\bf A}|)\Biggr]\ .\cr
&&\ 
\end{eqnarray}

	Expression (\ref{aqwe4}) exhibits a dependence on the trace $tr(V)$, which cannot be removed as before, in the masless case. So, the four-pole tensor $V$ cannot be defined as being traceless for the massive field.
	
	In order to compare the four-pole energies for the field with and without mass and write a single expression for both cases, let us proceed similarly to what we have done in the masless case and add a vanishing term to the energy (\ref{aqwe4}) given by
\begin{eqnarray}
\label{defDeltaE}
&&\!\!\!\!\!\!\! \Delta E_{III}(m,d)=\cr\cr
&&\!\!\!\!\!\!\! =\sigma\sigma_{0}\frac{tr V}{d}\Bigl({\bf\nabla}_{a\perp}^{2}-m^{2}\Bigr)\int\frac{d^{d}{\bf k}_{\perp}}{(2\pi)^{d}}\frac{1}{{\bf k}_{\perp}^{2}+m^{2}}\cr\cr
&&\ \ \ \ \ \ \ \exp{\bigl[i{\bf k}_{\perp}\cdot({\bf a}_{\perp}-{\bf A})\bigr]}\cr\cr
&&\!\!\!\!\! =-\frac{\sigma\sigma_{0}}{(2\pi)^{d/2}}m^{d/2}|{\bf a}_{\perp}-{\bf A}|^{-d/2}\cr\cr
&&\!\!\!\!\!\!\! \sum_{i,j=1}^{d}\frac{tr V}{d}\delta^{ij}\Biggl[\delta^{ij}K_{d/2}(m|{\bf a}_{\perp}-{\bf A}|)\cr\cr
&&\!\!\!\!\!\!\! -m|{\bf a}_{\perp}-{\bf A}|\frac{({\bf a}_{\perp}-{\bf A})^{i}}{|{\bf a}_{\perp}-{\bf A}|}\frac{({\bf a}_{\perp}-{\bf A})^{j}}{|{\bf a}_{\perp}-{\bf A}|}K_{(d/2)+1}(m|{\bf a}_{\perp}-{\bf A}|)\Biggr]\cr\cr
&&\!\!\!\!\!\!\! -\frac{\sigma\sigma_{0}}{(2\pi)^{d/2}}m^{d} \frac{tr V}{d}\ (m|{\bf a}_{\perp}-{\bf A}|)^{1-(d/2)}K_{(d/2)-1}(m|{\bf a}_{\perp}-{\bf A}|) .\cr
&\ &\!\!\!\!\!
\end{eqnarray}

	Using the Fourier representation for the Dirac delta function and the fact that ${\bf a}_{\perp}\not={\bf A}$, one can show that the right-hand side of the first line of Eq. (\ref{defDeltaE}) is equal to zero, so $\Delta E_{III}(m,d)=0$. Combining equations (\ref{aqwe4}) and (\ref{defDeltaE}), we get a new expression for the four-pole energy
\begin{eqnarray}
\label{tgb1}
&&\!\!\!\!\!\!\!\!\!\!\!\! E_{III}(m,d)\rightarrow E_{III}(m,d)+\Delta E_{III}(m,d)\cr\cr
&&\!\!\!\!\!\!\!\!\!\!\!\! =\frac{\sigma\sigma_{0}}{(2\pi)^{d/2}}m^{d/2}|{\bf a}_{\perp}-{\bf A}|^{-d/2}\cr\cr
&&\!\!\!\!\!\!\!\!\!\!\!\! \sum_{i,j=1}^{d}\Biggl[V^{ij}-\frac{tr V}{d}\delta^{ij}\Biggr] \Biggl[\delta^{ij}K_{d/2}(m|{\bf a}_{\perp}-{\bf A}|)\cr\cr
&&\!\!\!\!\!\!\!\!\!\!\!\! -m|{\bf a}_{\perp}-{\bf A}|\frac{({\bf a}_{\perp}-{\bf A})^{i}}{|{\bf a}_{\perp}-{\bf A}|}\frac{({\bf a}_{\perp}-{\bf A})^{j}}{|{\bf a}_{\perp}-{\bf A}|}K_{(d/2)+1}(m|{\bf a}_{\perp}-{\bf A}|)\Biggr]\cr\cr
&&\!\!\!\!\!\!\!\!\!\!\!\! -\frac{\sigma\sigma_{0}}{(2\pi)^{d/2}}m^{d} \frac{tr V}{d}\ (m|{\bf a}_{\perp}-{\bf A}|)^{1-(d/2)}K_{(d/2)-1}(m|{\bf a}_{\perp}-{\bf A}|)\ .\cr
&&\ 	
\end{eqnarray}

 By using the definition of the traceless tensor (\ref{defD}}), Eq. (\ref{tgb1}) can be rewritten in the form
\begin{eqnarray}
\label{aqwe5}
E_{III}(m,d)=
-\frac{\sigma_{0}}{(2\pi)^{d/2}}m^{1+(d/2)}|{\bf a}_{\perp}-{\bf A}|^{1-d/2}\cr\cr
\Biggl[\frac{tr(\sigma V)}{d}K_{(d/2)-1}(m|{\bf a}_{\perp}-{\bf A}|)\cr\cr
+\sum_{i,j=1}^{d}\!\!\!\frac{({\bf a}_{\perp}-{\bf A})^{i}}{|{\bf a}_{\perp}-{\bf A}|}(\sigma D^{ij})\frac{({\bf a}_{\perp}-{\bf A})^{j}}{|{\bf a}_{\perp}-{\bf A}|}K_{(d/2)+1}(m|{\bf a}_{\perp}-{\bf A}|)\Biggr] .\cr
\ \!\!
\end{eqnarray}

	Result (\ref{aqwe5}) is equivalent to (\ref{aqwe4}) and gives the interaction energy, intermediated by the massive scalar field, between a point charge placed at ${\bf a}$ and an uniform distribution of four-poles lying along a brane, placed at ${\bf A}$, and with four pole density $\sigma V$. In the limit $m\to0$, Eq. (\ref{aqwe5}) reduces to (\ref{aqwe2}). In (\ref{aqwe5}) we have, explicitly, separated the contribution to the energy due to the trace of the tensor $V$ and a contribution which does not come from the trace of $V$.
	
	For $d=3$, and considering a space-time with $3+1$ dimensions, we have a system composed by a point-like four-pole and a test charge, which has the corresponding interaction energy
\begin{eqnarray}
E_{III}(m,d=3)=-\frac{\sigma_{0}}{4\pi}m^{2}\frac{1}{|{\bf a}-{\bf A}|}\exp{(-m|{\bf a}-{\bf A}|)}\cr\cr
\Biggl[\frac{tr(\sigma V)}{3}+\frac{({\bf a}-{\bf A})^{i}}{|{\bf a}-{\bf A}|}(\sigma D^{ij})\frac{({\bf a}-{\bf A})^{j}}{|{\bf a}-{\bf A}|}\cr\cr
\times\Biggl(1+\frac{3}{m|{\bf a}-{\bf A}|}+\frac{3}{m^{2}|{\bf a}-{\bf A}|^{2}}\Biggr)\Biggr]\ .
\end{eqnarray}

	To conclude this section, we consider the stationary current distribution in $3+1$ dimensions given by
\begin{equation}
\label{ade1}
J_{n}({\bf x})=\sigma^{\mu_1\mu_2\mu_3...\mu_n}
[\partial_{\mu_1\mu_2\mu_3...\mu_n}^n\delta^{3}({\bf x}-{\bf A})]
+\sigma_{0}\delta^{3}({\bf x}-{\bf a}),
\end{equation}
that is, a point charge located at ${\bf a}$ and the derivative of arbitrary order of a Dirac's delta function concentrated at the point ${\bf A}$.
The quantity $\sigma^{\mu_1\mu_2\mu_3...\mu_n}$ is a completely symmetric tensor.
For the sake of simplicity, we consider only the four-dimensional case and point-like Dirac's delta functions.

	Replacing the current (\ref{ade1}) in (\ref{rfv1}) and proceeding as before, we obtain
\begin{equation}
\label{ade2}
E_{n}=-\frac{\sigma_0}{2}
\sigma^{\mu_1\mu_2\mu_3...\mu_n}
\partial_{({\bf a})\mu_1\mu_2\mu_3...\mu_n}^n\left(\frac{e^{-m|{\bf a}-{\bf A}|}}{4\pi |{\bf a}-{\bf A}|}\right)\ ,
\end{equation}
where $\partial_{({\bf a})\mu_1\mu_2\mu_3...\mu_n}^n$ means the derivative with respect to the ${\bf a}$ coordinates.

Considering the limit $m\to 0$ and defining the vector ${\bf r}={\bf a}-{\bf A}$ we get
\begin{equation}
\label{ade3}
E_n=-\frac{\sigma_0}{8\pi}\sigma^{\mu_1\mu_2\mu_3...\mu_n}
\partial_{\mu_{1}\mu_{2}\mu_{3}...\mu_{n}}^n\frac{1}{r}.
\end{equation}

	It can be easily verified that the expression above gives the interaction between a point charge and an $N$-pole. For instance, whenever $n=3$, we have the interaction between a point charge and a four-pole; for $n=4$, the interaction between a point charge and an octupole, and so on.

\section{Maxwell Field}
\setcounter{equation}{0}
\label{eletromagnetico}

	In this section, we extend the results obteined for the scalar field, in the previous section, for the electromagnecic field. We always take models described by the lagrangian density
\begin{equation}
\label{Leletromagnetico}
{\cal L}=-\frac{1}{4}F_{\mu\nu}F^{\mu\nu}-\frac{1}{2\alpha}(\partial_{\mu}A^{\mu})^{2}+J^{\mu}A_{\mu}\ ,
\end{equation}
where $A^{\mu}$ is the electromagnetic field, $F_{\mu\nu}=\partial_{\mu}A_{\nu}-\partial_{\nu}A_{\mu}$ is the field strength and $J^{\mu}$ is a stationary external current, different for each model we consider. In the above equation $\alpha$ is a gauge parameter.

	Following similar steps which lead to equation (\ref{rfv1}), the vacuum energy corresponding to the lagrangian (\ref{Leletromagnetico}) can be writen in the form
\begin{equation}
\label{arfv1}
E=\frac{1}{2T}\int\int d^{d+D+1}x d^{d+D+1}x\ J^{\mu}(x)\ \Delta_{\mu\nu}(x,y)\ J^{\nu}(y)\ ,
\end{equation}
where $\Delta_{\mu\nu}(x,y)$ is the photon propagator
\begin{eqnarray}
\label{defDeltamunu}
&&\!\!\!\!\! \Delta_{\mu\nu}(x,y)=\nonumber \\
&&\!\!\!\!\! -\int\frac{d^{d+D+1}k}{(2\pi)^{d+D+1}}\frac{1}{k^{2}}\Biggl[\eta_{\mu\nu}-(1-\alpha)\frac{k_{\mu}k_{\nu}}{k^{2}}\Biggl]\exp{[-ik(x-y)]}\ .\cr
&&\ 
\end{eqnarray}
	
	The first model we study is given by the current
\begin{equation}
\label{corrente4}
J_{IV}^{\mu}=\sigma W^{\mu} \delta^{d}({\bf x}_{\perp}-{\bf A})+\sigma_{0} \eta^{\mu 0} \delta^{d+D}({\bf x}-{\bf a})\ ,
\end{equation}
which is a generalization of the one specified by the current (\ref{corrente1}). The quantity $\ W^{\mu}$ is a $(d+D+1)$-vector taken to be constant and uniform in the reference frame we are performing the calculations. In order to ensure gauge invariance for the last term in the right-hand side of the lagrangian (\ref{Leletromagnetico}), the $(d+D+1)$-vector $W^{\mu}$ must satisfy the condition ${\bf W}_{\perp}=0$.

	Inserting (\ref{corrente4}) into (\ref{arfv1}), using the Fourier representation (\ref{defDeltamunu}), performing, in the following order, the integrals $dx^{0}dk^{0}dy^{0}d^{D}{\bf x}_{\|}d^{D}{\bf y}_{\|}d^{D}{\bf k}_{\|}d^{d}{\bf x}_{\perp}d^{d}{\bf y}_{\perp}$ and taking into account that ${\bf W}_{\perp}=0$ we have
\begin{equation}
E_{IV}(d)=\sigma_{0}(\sigma W_{0})\int\frac{d^{d}{\bf k}_{\perp}}{(2\pi)^{d}}\frac{1}{{\bf k}_{\perp}^{2}}\exp{[i{\bf k}_{\perp}\cdot({\bf a}_{\perp}-{\bf A})]}\ .
\end{equation}
In reference \cite{BaroneHidalgo}, the above integral is calculated for any $d\not=2$,
\begin{equation}
\label{xsw1}
E_{IV}(d)=\frac{\sigma_{0}(\sigma W_{0})}{(2\pi)^{d/2}}\Gamma\bigl((d/2)-1\bigr)2^{(d/2)-2}|{\bf a}_{\perp}-{\bf A}|^{2-d}\ ,\ d\not=2\ .
\end{equation}

	If we take $W^{\mu}\sim\eta^{\mu0}$, the result (\ref{xsw1}) becomes the interaction energy between a point-like test charge placed at ${\bf a}$ and a uniform charge distribution along a $D$-brane with charge density $\sigma$ and placed at ${\bf A}$.

	For $d=2$, we insert a mass parameter in the propagator (\ref{defDeltamunu}) as a regulator parameter in order to identify infrared divergences. This procedure leads to the expression
\begin{eqnarray}
E_{IV}(d=2)&=&\lim_{m\to 0}\sigma_{0}(\sigma W_{0})\int\frac{d^{d}{\bf k}_{\perp}}{(2\pi)^{d}}\frac{1}{{\bf k}_{\perp}^{2}+m^{2}}\nonumber \\
&&\exp{[i{\bf k}_{\perp}\cdot({\bf a}_{\perp}-{\bf A})]}\cr\cr
&=&\lim_{m\to 0}\frac{\sigma_{0}(\sigma W_{0})}{2\pi}K_{0}(m|{\bf a}_{\perp}-{\bf A}|)\cr\cr
&\to&-\frac{\sigma_{0}(\sigma W_{0})}{2\pi}\ln\Biggl(\frac{|{\bf a}_{\perp}-{\bf A}|}{a_{0}}\Biggr)\ ,
\end{eqnarray}
where we have proceeded similarly to what we have done in Eq. (\ref{qwe4}).

	The second model we study for the Maxwell field is given by the current
\begin{equation}
\label{corrente5}
J_{V}^{\mu}=\sigma W^{\mu} V^{\alpha}\partial_{\alpha}\bigl(\delta^{d}({\bf x}_{\perp}-{\bf A})\bigr)+\sigma_{0} \eta^{\mu 0} \delta^{d+D}({\bf x}-{\bf a})\ ,
\end{equation}
where $W^{\mu}$ and $V^{\mu}$ are four-vectors defined in the same way as in Eq's (\ref{corrente4}) and (\ref{corrente2}), respectively.

	Inserting the current (\ref{corrente5}) in the energy (\ref{arfv1}), using definition (\ref{defDeltamunu}) and performing similar steps which lead to the result (\ref{azxc2}) from (\ref{rfv1}) we have
\begin{eqnarray}
\label{EV}
E_{V}(d)&=&\frac{\sigma_{0}}{(2\pi)^{d/2}}\Gamma(d/2)2^{(d/2)-1}\cr\cr
&\ &|{\bf a}_{\perp}-{\bf A}|^{-d}(-\sigma W^{0}{\bf V})\cdot({\bf a}_{\perp}-{\bf A})\ .
\end{eqnarray}

	If we take $W^{\mu}\sim\eta^{\mu 0}$, Eq. (\ref{EV}) can be interpreted as the interaction energy between a point-like charge and a uniform distribution of electric dipoles along a $D$-brane, with dipole density given by $-\sigma V^{\alpha}$. In order to make this fact clearer let us take $d=3$, $W^{\mu}=\eta^{\mu0}$ and restrict ourselves to a $3+1$ spacetime. In this case, we have a point-like dipole and the energy (\ref{EV}) reads \cite{Jackson,Landau}
\begin{equation}
E_{V}(d=3)=\frac{\sigma_{0}}{4\pi}\frac{(-\sigma{\bf V})\cdot({\bf a}-{\bf A})}{|{\bf a}-{\bf A}|^{3}}\ .
\end{equation}

  The third and last model we consider for the Maxwell field is described by the current
\begin{equation}
\label{corrente6}
J_{VI}^{\mu}=\sigma W^{\mu}V^{\alpha\beta}\partial_{\alpha}\partial_{\beta}\bigl(\delta^{d}({\bf x}_{\perp}-{\bf A})\bigr)+\sigma_{0}\eta^{\mu\nu}\delta^{d+D}({\bf x}-{\bf a})\ ,
\end{equation}
where $V^{\alpha\beta}$ is a tensor with the same features of the one used in the current (\ref{corrente3}) and $W^{\mu}$ is the same $(d+D+1)$-vector used in (\ref{corrente4}).

	Inserting the current (\ref{corrente6}) in the expression (\ref{arfv1}) and performing similar steps which lead to the result (\ref{aqwe2}), we have, for the electromagnetic field, the energy
\begin{eqnarray}
\label{EVI}
E_{VI}(d)&=&\frac{\sigma_{0}}{\pi^{d/2}}\Gamma\bigl((d/2)+1\bigr)|{\bf a}_{\perp}-{\bf A}|^{-d-2}\cr\cr
&\ &\sum_{i,j=1}^{d}({\bf a}_{\perp}-{\bf A})^{i}(\sigma W^{0}D^{ij})({\bf a}_{\perp}-{\bf A})^{j} ,
\end{eqnarray}
where we have used definition (\ref{defD}).

	For $W^{\mu}\sim\eta^{\mu0}$, we can interpret Eq. (\ref{EVI}) as the interaction energy between a point-charge and an uniform four-pole distribution along a $D$-brane, with four-pole density given by $\sigma D^{ij}$. In $3+1$ dimensions with $W^{\mu}=\eta^{\mu0}$ the energy (\ref{EVI}) reads
\begin{equation}
E_{VI}(d=3)=\frac{3\sigma_{0}}{4\pi}|{\bf a}_{\perp}-{\bf A}|^{-5}\sum_{i,j=1}^{d}({\bf a}-{\bf A})^{i}(\sigma D^{ij})({\bf a}-{\bf A})^{j}\ ,
\end{equation}
which is the interaction energy between a point-like four-pole $(\sigma D^{ij})$ placed at ${\bf A}$ and a test charge $\sigma_{0}$ placed at ${\bf a}$ \cite{Jackson,Landau}.

	To conclude this section, we stress that we could obtain, for the electromagnetic field, a result similar to the one presented in equation (\ref{ade3}) for $n$-pole distributions.

\section{Conclusions and Final Remarks}
\label{conclusao}

	In this paper, we have carried out an investigation on the role of external currents concentrated at specific regions of space ($D$-dimensional branes) and coupled to bosonic fields, specifically, the scalar and electromagnetic ones.
	
	We have considered a $D+d+1$ dimensional space-time and three kinds of currents for each field. All currents are composed by two parts where the second one describe the presence of a stationary point-like test-charge used to investigate the force field produced by the former one. From the results obtained for masless fields, we could notice that the first term of each current describe the presence of stationary charges, dipoles or four-poles distributions along $D$-dimensional branes.

	As for the results for dipole-distributions (the second and fifth models studied), we corrected a flaw in reference \cite{BaroneHidalgo} on the interpretation of the dipole-distribution sign described by the currents.

	We have shown that, for masless fields, the four-pole tensor density corresponding to a uniform four-pole distribution along a $D$-brane can be defined as being traceless in any dimension, once its trace does not contribute to the interaction energy between a test charge and the corresponding four-pole distribution.

	For massive fields, we have shown that the trace of the four-pole tensor density always contributes to the interaction energy between a point-like test charge and the corresponding four-pole distribution. This fact can be seen from Eq. (\ref{aqwe5}), where we have the interaction energy between a point-like test-charge and a four-pole distribution mediated by the massive scalar field. In this result, we have separated the contributions to the energy due to the traceless part of the four-pole density tensor, $\sigma D$, and the contribution due strictly to the trace of this tensor, $\sigma\ tr D$.
	
Finally, we have obtained that, if we consider a massless scalar field in four-dimensional space-time in interaction with a stationary current composed by a point-like Dirac's delta function and the $n$-th derivative of a point-like Dirac's delta function, the result is the interaction energy between a point charge and a point-like $N$-pole. We stress that an identical result can be obtained also in the electromagnetic case.

	\

{\bf Acknowledgements}

	The authors would like to thank C. Farina, J.A. Helay\"el-Neto and N.F Svaiter for discussions and suggestions, J.A. Helay\"el-Neto and F.E. Barone for reading the manuscript and FAPEMIG for invaluable financial support.



\end{document}